\documentclass[preprint,aps,floatfix,
preprintnumbers,pre]{revtex4}
\usepackage{amsmath}
\usepackage{amssymb}
\usepackage{epsfig}
\usepackage{graphicx}
\usepackage{dcolumn}
\usepackage{bm}
\newcommand{\tmop}[1]{\operatorname{#1}}

\newcommand{\mathd}{\mathrm{d}}
\newcommand{\mathe}{\mathrm{e}}
\newcommand{\tmmathbf}[1]{\boldsymbol{#1}}

\begin{document}

\title{Quantum Adiabatic Evolution Algorithm and Quantum Phase Transition
in 3-Satisfiability Problem}
\author{S. Knysh$^{1,2}$ and V. N. Smelyanskiy$^{1}$}
\affiliation{
$^{1}$NASA Ames Research Center, Moffett Field, CA 94035\\
$^{2}$Mission Critical Technologies, El Segundo, CA 90245}

\date{\today}

\begin{abstract}
In this paper we show that the performance of the quantum
adiabatic algorithm is determined by phase transitions in
underlying problem in the presence of transverse magnetic field
$\Gamma$. We show that the quantum version of random
Satisfiability problem with 3 bits in a clause (3-SAT)  has a
first-order quantum phase transition. We analyze the phase diagram
$\gamma=\gamma(\Gamma)$ where $\gamma$ is an average number of
clauses per binary variable in 3-SAT. The results are obtained in
a closed form assuming replica symmetry and neglecting time
correlations at small values of the transverse field $\Gamma$. In
the limit of $\Gamma=0$ the value of $\gamma(0)\approx$ 5.18
corresponds to that given by the replica symmetric treatment of a
classical random 3-SAT problem. We demonstrate the qualitative
similarity between classical and quantum versions of this problem.
\end{abstract}

\maketitle

\section{Introduction}

From the early years of computing it was realized that some
problems are inherently intractable. This intuition has been
quantified by the theory of computational complexity proposed by
Cook \cite{cook}. Most problems of practical interest can be
roughly divided into two classes: P and NP-complete. Problems in
the former class can be solved on a computer in time that scales
only polynomially with the size of an instance of the problem.
Solution to NP-complete problems can be {\em verified} in
polynomial time, but it is believed that it cannot be found in
polynomial time on a classical computer. Solving NP-complete
problem typically requires exponential time which makes then
intractable. All NP-complete problems are equivalent; if it were
possible to solve one NP-complete problem in polynomial time, it
would be possible to apply the same algorithm to solve {\em all}
NP-complete problems.

It is not known yet whether NP-complete problems can be solved efficiently on a
quantum computer. Shor's algorithm \cite{shor}
for a quantum computer can solve in
polynomial time the number factoring problem that is presumably hard for a
classical computer. It is not, however, in the NP-complete class (and its
decision variant is in P).

Just as research in classical computing focus on worst-case
complexity has been superseded by the analysis of typical case
complexity and application of general-purpose algorithms like
simulated annealing, similar changes take place in the field of
quantum computing. A substantial interest has been generated by a
general-purpose quantum adiabatic  algorithm (QAA) proposed by
Farhi and coworkers \cite{farhi,FarhiSc}.

In its simplest form QAA is applied to problems where underlying
variables can have only two values, and a solution is given by a
N-bit binary string. It corresponds to a time-dependent
Hamiltonian that slowly changes from a simple form (for which the
ground state can be constructed exactly) to a complex form that
describes an instance of NP-complete problem. The initial
Hamiltonian is typically chosen to correspond to a uniform
magnetic field $\Gamma$ applied along $\hat{x}$ direction
\begin{equation}
   H_\textrm{in} = - \Gamma \sum_{i=1}^{N}
   \hat{\sigma}_i^x,\label{Hin}
\end{equation}
so that the ground state in $\{ \hat{\sigma}_i^z \}$ basis is a
symmetric superposition of $2^N$ binary vectors
\begin{equation}
   \psi_\textrm{in} = \frac{1}{2^{N / 2}}  \sum_{\{ s_i=\pm 1 \}} |s_1 s_2 \cdots
   s_N \rangle .\label{psiin}
\end{equation}
The final Hamiltonian is chosen to be diagonal in $\{
\hat{\sigma}_i^z \}$ basis with diagonal elements corresponding to
the cost function values for particular assignments of binary
variables $\{ s_i \}$
\begin{equation}
   H_{\rm fin} = \sum_{\{ s_i \}} E [ \{ s_i \} ] |s_1 s_2 \cdots s_N
   \rangle \langle s_1 s_2 \cdots s_N |.\label{Hfin}
\end{equation}
At intermediate times the Hamiltonian is a linear combination of initial and
final Hamiltonians. We choose to write this in the following form
\begin{equation}
   H = \sum_{\{ s_i \}} E [ \{ s_i \} ] |s_1 \cdots s_N \rangle \langle s_1
   \cdots s_N | - \Gamma \sum_i \widehat{\sigma_{}}_i^x .\label{H}
\end{equation}
In the beginning of the algorithm $\Gamma \rightarrow \infty$ so
that the second term dominates and a ground state has a simple
form. At the end of the algorithm $\Gamma = 0$ and the ground
state corresponds to a solution of the NP-complete problem that
minimizes cost function. If $\Gamma$ is lowered sufficiently
slowly, adiabatic theorem tells us that the system will remain in
its ground state with a high probability. The second term is
referred to as a {\em driver} term because its presence in
otherwise diagonal Hamiltonian is to enable spin flips. The
algorithm is quite similar to the simulated annealing (SA)
algorithm in this respect. In QAA transverse field $\Gamma$
replaces temperature $T$. The dynamics of these two problems are
quite different.

\begin{figure}[!ht]
\includegraphics[scale=0.67]{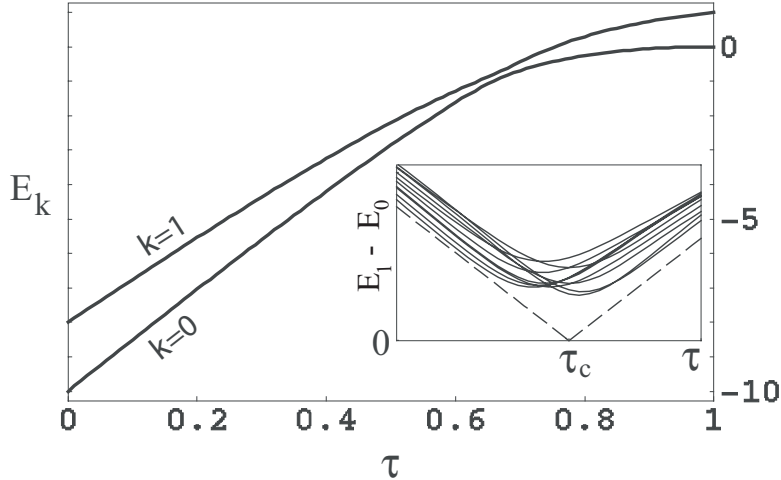}
\caption{Two lowest eigenvalues of the  Hamiltonian $H$ (\ref{H})
{\it vs} the interpolating parameter $\tau=1/(1+\Gamma)$. In
simulations $H_{\rm fin}$ corresponds to a randomly generated
instance of the version of Satisfiability problem  Exact Cover
\cite{FarhiSc} with the number of bits $N$=10. {\it Insert}:
different plots show the $\tau$-dependencies of the gap between
the two lowest eigenvalues of $H$ for various random instances of
 Exact Cover with $N$=10.  Dashed lines indicate the
approximate asymptotical position of the avoided-crossing point in
the limit $N\rightarrow\infty$. \label{fig:ac}}
\end{figure}

The fundamental limitation of the SA algorithm is the critical
slowing down at the point of classical phase transition. The
system may become trapped in one of the  local minima surrounded
by high barriers and that requires exponentially long time to
escape. For the QAA the metric of the algorithm performance can be
given in terms of the eigenstates $\Psi_k(\Gamma)$ and eigenvalues
$E_k(\Gamma)$ of the interpolating Hamiltonian $H(\Gamma)$. The
rate at which $\Gamma(t)$ can be lowered obeys the strong
inequality $\hbar |\dot\Gamma(t)|\langle \Psi_0|
\,dH/d\Gamma|\Psi_0\rangle \ll (E_1(\Gamma)-E_0(\Gamma))^2$. A
sufficient condition for the success of QAA requires a polynomial
scaling with the problem size $N$ of a minimum gap $\Delta E_{\rm
min}=\min_{\,\Gamma\in(0,\infty)}\left(E_1(\Gamma)-E_0(\Gamma)\right)$
 between the two lowest energy levels of $H(\Gamma)$.
 The danger is
that there are points of {\em avoided crossing} ((see
Fig.~\ref{fig:ac})) at which the gap can be exponentially small.
\noindent
 At such point the branches of $E_0(\Gamma)$ correspond to wavefunctions with
exponentially small overlap. This effect has been demonstrated
analytically in a toy problem \cite{Vazirani02,qaetoy} as well as
for the case of NP-complete problem Positive-1-in-K-sat with large
number of bits in a clause, $K \gg 1$ \cite{skm:04}.

An equivalent measure of the algorithmic complexity is the second
derivative of the ground state energy $\mathd^2 E_0 / \mathd
\Gamma^2$. This quantity can
 become exponentially large at
the point of avoided crossing where it is $\propto 1/\Delta E_{\rm
min}$. If the minimum gap $\Delta E_{\rm min}$ shrinks to zero in
the thermodynamic limit $N \rightarrow \infty$ for some
$\Gamma=\Gamma_c$ then the ground state energy $E_0(\Gamma)$
possesses a cusp (cf. Fig.~\ref{fig:ac}). This cusp signals a
first-order quantum phase transition. Indeed, the ground state
energy is precisely the free energy in the limit $\beta = 1 / T
\rightarrow \infty$:
\begin{equation} E_0 ( \Gamma ) \equiv \lim_{\beta \rightarrow \infty} F ( \beta, \Gamma ) =
   - \lim_{\beta \rightarrow \infty} \frac{1}{\beta} \ln \textrm{Tr} \left\{
   e^{- \beta \hat{H} ( \Gamma )} \right\} . \end{equation}
Time evolution has to be slowest at the point of the first-order
quantum phase transition, reminiscent of a critical slowing down
at the point of classical phase transition.

We will be working with an ensemble of random instances of
NP-complete problem and recast it as essentially a spin glass
model. Although quantum phase transitions in infinitely-connected
models of spin glasses have been extensively studied
\cite{qsg1,dynamic,dynamic2,dynamic3}, there are no results to our
knowledge on dilute quantum spin glasses. At the same time most of
NP-complete problems are closely related to dilute spin glass
models.

The central quantity we will be computing is the disorder-averaged
(instance-averaged) free energy $\langle F ( \beta ) \rangle$. We
will classify the quantum phase transition as random first-order
or random second-order as deduced from the non-analyticity in
$\langle F \rangle$ {\it vs} $\Gamma$. In classical case, random
first-order phase transition is typically associated with
exponentially hard problems. It is therefore tempting to suggest
that the onset of the first-order quantum phase transition is
indicative of the exponential, or stretched-exponential
\cite{Fisher,Reichardt} scaling law of the runtime of QAA with $N$
determined by the distribution of the tunnelling times between the
valleys of an energy landscape.

Within current formalism we are limited to this qualitative
picture. The determination of the minimum gap requires more
elaborate methods. One would naively expect that finite-size
scaling analysis can be used to derive the scaling of the minimum
gap with $N$.
 Observe though that, if disorder is relevant and
 assuming the first-order phase transition, the position of the minimum gap
fluctuates with disorder around some value $\Gamma_c$ (see
Fig.~\ref{fig:ac}). If we fix a value of $\Gamma$ and perform the
disorder-average, we will have $\langle E_1(\Gamma) -
E_0(\Gamma)\rangle \gtrsim O ( 1 / \sqrt{N} )$. Minimized over
$\Gamma$, it is still $O ( 1 / \sqrt{N} )$ even though true
minimum gap can be arbitrarily small.

The paper is organized as follows. In Sec.~\ref{sec:ksat} we
introduce the random K-SAT problem -- the model that we study in
this paper. In Sec.~\ref{sec:review} we describe exact results
from the random graph theory and empirical results on
satisfiability transition. K-SAT problem has been extensively
studied in the classical limit for zero temperature. This section
is devoted to these results as well as effects of finite
temperature. The Sec.~\ref{sec:q3sat} formulates the K-SAT problem
in quantum case in the presence of transverse magnetic field.
Subsequently two approximations are made to make the problem
tractable: the so-called replica-symmetric ansatz that assumes
absence of long-range correlations is made, and the static ansatz
that assumes that any correlations time correlations are due to
the transverse magnetic field alone. The expression for the free
energy and the self-consistency equation for the order parameter
are derived. In the Sec.~\ref{sec:small} we consider the limit of
small transverse magnetic field $\Gamma$ and derive the simplified
self-consistency equations for this case.
Sec.~\ref{sec:conclusion} is Conclusion.

\section{\label{sec:ksat}K-SAT Problem}

As a test case for QAA algorithm we consider the satisfiability
problem that happens to be the first problem to be associated with
NP-complete class \cite{cook}. More precisely, we will work with a
variant of satisfiability problem -- the K-SAT that places a
constraint on the number of variables that can appear in a clause.
The K-SAT is known to be NP-complete for $K \geqslant 3$. This
means that any NP-complete problem can be reformulated as an
instance of 3-SAT. We will concentrate on this particular case ($K
= 3$). The benefits of working with K-SAT is that the problem
Hamiltonian is local (K-local to be exact, but can easily be
recast as a 2-local Hamiltonian; see \cite{qaesolid} for
application to QAA algorithm); and also that random K-SAT is
intrinsically related to random hypergraphs and is amenable to
methods of statistical physics.

An instance of K-SAT is a boolean formula in conjunctive normal form (CNF),
i.e. a set of clauses
\begin{equation}
   \mathcal{F}=\mathcal{C}_1 \wedge \mathcal{C}_2 \wedge \cdots \wedge
   \mathcal{C}_M
\end{equation}
where each clause represents a logical OR of $K$ literals, each literal
representing either one of the variables $x_1, \ldots, x_N$ or a its logical
NOT. Below is an example of possible K-SAT clause for $K = 3$:
\begin{equation}
   x_2 \vee \bar{x}_5 \vee \bar{x}_{11}
\end{equation}
(where $\bar{x}$ represents logical NOT of $x$).

The formula is said to be satisfiable if and only if there exists an
assignment of boolean variables $\{ x_i \}$ such that all $M$ clauses
comprising a formula are satisfied (evaluate to {\em true}) at the same
time. Presented with an instance of K-SAT formula, algorithm must determine if
it is satisfiable and if it is, find the appropriate assignment of variables.

We follow the standard recipe for trivial mapping of constraint satisfaction
problems to problems in statistical mechanics by defining the energy function
to be proportional to the number of violated clauses. The constant of
proportionality is chosen to be $2$. The motivation for this choice will
become evident later. The energy can be written as follows
\begin{equation}
   E = 2 \sum_m \theta ( J_{m 1} s_{i_{m 1}} ) \theta ( J_{m 2} s_{i_{m 2}} )
   \cdots \theta ( J_{m K} s_{i_{m K}} ) . \label{eq:H-KSAT}
\end{equation}
where we replaced boolean variables with spins $s_i = \pm 1$, the value $s = -
1$ corresponding to {\em true}. $\theta ( x )$ is the Heaviside function
$\theta ( x ) = 1$ if $x > 0$ and $\theta ( x ) = 0$ if $x \leqslant 0$.
Variables $i_{m 1}, \ldots ., i_{m K}$ are indices of variables that appear in
a clause, and $J_{m 1}, \ldots, J_{m K} = \pm 1$ describe whether the literal
that appears in a clause is negated ($J = - 1$).

We will work with an ensemble of random instances of K-SAT with fixed number
of variables $N$, number of clauses $M$ and number variables per clause $K$.
This merely means that indices $i_{m p}$ are independent and drawn uniformly
at random from the set $1, 2, \ldots, N$ and variables $J_{m p}$ are
independent random variables that are equal to $+ 1$ or $- 1$ with probability
50 \% .

The formula is satisfiable if the ground state energy equals zero, and it is
unsatisfiable if it is greater than zero.

The following phenomenon has been discovered \cite{satunsat}.
If we fix the ratio of clauses
to variables $\gamma = M / N$, the energy is almost surely $0$ for $\gamma <
\gamma_c$ in the limit $N \rightarrow \infty$ and $E > 0$ for $\gamma >
\gamma_c$. The value of $\gamma_c$ is independent of $N$ in the limit of large
$N$, apart from small corrections. Note that the ground state remains
exponentially degenerate across the transition. The exact value of $\gamma_c$
is an interesting problem in combinatorics \cite{mathconst};
rigorous determination of $\gamma_c$ has not yet been accomplished.

The threshold phenomena in random graphs were known since pioneering work of
Erdos and Renyi \cite{erdos}.
Random K-SAT problem has intimate connections to the theory
of random graphs. One can picture an instance of random K-SAT as a random
hypergraph. Variables correspond to vertices of the hypergraph and clauses
correspond to hyperedges that join $K$ vertices. One threshold phenomenon that
was examined in \cite{erdos} in the context of random graphs but can be
extended to random hypergraphs, is the appearance of giant component.
In physics it is known as percolation transition.
When $\gamma < 1 / K ( K - 1 )$ a random
hypergraph is with high probability is a collection of disjoint clusters of
typical size $O ( 1 )$, with the largest cluster not exceeding $O ( \log N )$
in size. All clusters (with few (i.e. $O ( 1 )$) exceptions) have a form of
trees, that is contain no loops. For $\gamma > 1 / K ( K - 1 )$ the same
picture holds, but there appears a giant (of size $O ( N )$) supercluster. The
structure of the giant supercluster is very complex. It contains a large
number of loops, but the length of these loops is relatively large: $O ( \log
N )$. It is in the giant supercluster that the complexity of the problem is
buried. Indeed, small isolated clusters can be efficiently tackled using
divide-and-conquer approach. Moreover, since they are all tree-like, the
assignment of variables that satisfies corresponding clauses always exists and
can be found in linear time.

\begin{figure}[!ht]
\includegraphics[scale=0.67]{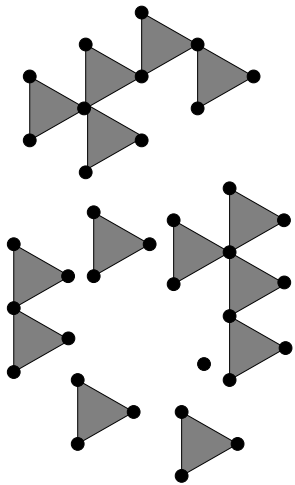}
\caption{Example of hypergraph corresponding to random instance of 3-SAT.
Vertices represent variables and traingles represent clauses involving $K=3$
variables.}
\end{figure}

Although phase transition is evident in geometric properties, owing to long
length of the loops in the supercluster, no abnormal behavior in the space of
solutions is seen immediately following the phase transition. The formula
remains satisfiable well past the percolation transition, nor any correlations
between spins in the supercluster that are far away from each other are
detected. One must note that such correlations do appear at so called dynamic
transition $\gamma_d$, at which point finding the solution becomes difficult
(for simulated annealing algorithm).

\begin{figure}[!ht]
\includegraphics{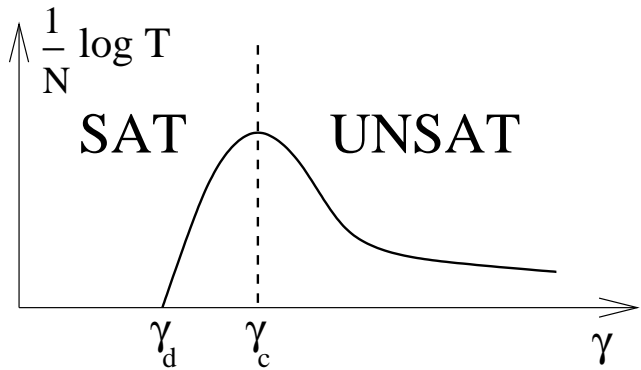}
\caption{Complexity of the instance as the function of $\gamma=M/N$. Problem
becomes exponentially hard for $\gamma > \gamma_d$ and the complexity peaks
at satisfiability threshold $\gamma_c$.}
\end{figure}

The performance of algorithm is subject to similar threshold phenomenon.
Typically, algorithms require just a polynomial (often linear) time for
$\gamma$ less than some critical value $\gamma_d$ and require an exponential
time for $\gamma > \gamma_d$. One can introduce the normalized complexity
$\lim_{N \rightarrow \infty} \frac{1}{N} \log T$, where $T$ is the time it
takes to solve a problem (units of $T$ are unimportant in $N \rightarrow
\infty$ limit since the logarithm is taken). It is a self-averaging quantity
and hence a function of $\gamma$ alone (in the usual sense, with high
probability a random formula has this complexity). This function equals zero
for $\gamma < \gamma_d$ and for $\gamma > \gamma_d$ it is non-zero. The exact
value of $\gamma_d$ and the shape of the complexity function is not universal,
but rather algorithm-dependent. Simulations suggests that for most common
algorithm the complexity peaks at $\gamma_c$ that is the point SAT/UNSAT
transition. Indeed, the satisfiability threshold is the place where one
intuitively expects to find the hardest (and most interesting) problems. The
exact determination of the complexity function is invaluable as a
non-empirical means for comparing the asymptotic efficiency of various
algorithm. This is particularly useful in the field of quantum computing,
where empirical research is impossible since no prototype of quantum computer
exist and its simulation on classical computer is not feasible for large
instances of the problem. An important benchmark of the algorithm is
$\gamma_d$ itself (larger is better) as it marks a region where the problem
can be solved very efficiently. An alluring (but not necessarily impossible)
goal is designing an algorithm that has $\gamma_d \geqslant \gamma_c$.

Although the value of $\gamma_d$ is algorithm-dependent, a particular value
that we denote $\gamma_\textrm{RSB}$ is universal in a sense that it should be
the same for all local search algorithms. It corresponds to the point where
the energy landscapes qualitatively changes. There appears an exponentially
large number of local minima, i.e. a set of states corresponding to $E = 0$
becomes disconnected. In contrast for $\gamma < \gamma_\textrm{RSB}$ a set of
states with $E = 0$ is connected. This transition can be seen in the behavior
of the free energy. This is the point where so-called replica-symmetric ansatz
breaks down. This will be elaborated on in the next section.

\section{\label{sec:review} Review of Classical Result}

In a number of articles \cite{ksat1,ksat2,ksat3,ksat4,ksat5} satisfiability
transition has been studied in
K-SAT problem using replica method. Statistical properties of physical system
are completely determined by its free energy
\begin{equation}
   F = - T \ln Z = - T \ln \sum_{\{ s_i \}} \mathe^{- \beta H [ \{ s_i \} ]} .
\end{equation}
If the Hamiltonian depends on the disorder, we must also average the free
energy over disorder configurations. Note that it is not appropriate to
attempt to compute the disorder average of $Z$. Indeed $Z$ is exponentially
large and small $o ( N )$ fluctuations of $F$ amplify in $Z = \mathe^{- \beta
F}$. Therefore
\begin{equation}
   \langle Z \rangle = \left\langle \mathe^{- \beta F}
   \right\rangle \neq \mathe^{- \beta \langle F \rangle} .
\end{equation}
Note that although the free energy $F$ is $O ( N )$, fluctuations due to
disorder are only $O ( \sqrt{N} )$ and can be neglected in the thermodynamic
limit. Disorder averaging of $F$ is just a useful trick, since to the leading
order in $N$, same value should be obtained for $F$ for almost all possible
disorder realizations (all but a fraction that goes to $0$ with increasing
$N$).

Replica method accomplishes the averaging as follows. $n$ non-interacting
copies of the system are prepared, all having the identical disorder
configuration. This is indicated by attaching an additional replica index
$\alpha = 1, \ldots, n$ to each variable $s_i$; new variables are labeled
$s_i^{\alpha}$. The partition function of such system equals $Z^n$, where $Z$
is the partition function of non-replicated system. Disorder average of the
replicated system is then performed, which is easy to accomplish since
summation over all possible spin configurations $\{ s_i^{\alpha} \}$ can be
done after the disorder averaging, which eliminates the difficulty of
computing the partition function that explicitly depends on the disorder. Once
analytical expression for $\langle Z^n \rangle$ is obtained, the
disorder-averaged free energy $\langle F \rangle = - T \langle \ln Z \rangle$
is computed via analytical continuation using the identity
\begin{equation}
   \langle \ln Z \rangle = \left. \frac{\mathd}{\mathd n} \langle Z^n \rangle
   \right|_{n = 0}
\end{equation}
The Hamiltonian for the K-SAT problem has been introduce in the previous
section (Eq. \ref{eq:H-KSAT}).
In the limit of zero temperature satisfiable phase is
characterized by $\langle E \rangle = 0$; in the unsatisfiable (UNSAT) phase
$\langle E \rangle > 0$.

In the simplest approximation it is assumed that the symmetry of the
Hamiltonian with respect to the interchange of replica indices $\alpha$ is not
spontaneously broken. The physical interpretation of this is the absence of
long-range correlations. For randomly chosen variables $s_i$ and $s_j$,
replica symmetric ansatz implies
\begin{equation}
   \langle s_i s_j \rangle = \langle s_i \rangle \langle s_j \rangle
   \label{eq:rs}
\end{equation}
(randomly chosen sites are with high probability at least $O ( \log N )$ away
from each other; correlations are absent in the limit $N \rightarrow \infty$).

Owing to identity (\ref{eq:rs}) statistical properties (correlations) in the
thermodynamic limit are completely determined by specifying average
magnetizations of individual spins $m_i = \langle s_i \rangle$. The order
parameter of the system is the histogram of magnetizations of spins $P ( m )$.
In practice it is convenient to define effective fields $h_i = T \tanh^{- 1}
m_i$ and choose the histogram of effective fields as an order parameter. An
expression for the free energy can be written out in terms of $P ( h )$.
\begin{eqnarray}
   F & = & \frac{1}{\beta} \int \mathd h \ln(2 \cosh \beta h)
           \int \frac{\mathd \omega}{2\pi} \tilde{P}(\omega)
           (\ln \tilde{P}(\omega)-1) \nonumber \\
   & & -\frac{\gamma}{\beta} \int \prod_{i=1,2,3} \mathd h_i P(h_i)
      \ln \left[ 1-\frac{1-\mathe^{-\beta}}
                  {\prod_{i=1,2,3}(1+\mathe^{2\beta h_i})} \right]
   \label{eq:F}
\end{eqnarray}
Varying this expression with respects to $P ( h )$ yields a self-consistency
equation of $P ( h )$ which can be solved numerically. This equation is
essentially an iterative procedure for determining correct magnetic fields for
clusters without loops (trees) combined with the Poisson distribution for the
number of branches in a random tree.

For high connectivities $\gamma$ and low temperatures $T$ solution to the
self-consistency equation yields two solutions. A solution that
{\em maximizes} the free energy should be chosen. Note that this is in
contrast to standard of choosing a solution with the smaller free energy for
first-order phase transition in pure systems. This reversal is standard
feature of replica method and the rationale is discussed in \cite{sgbook}.

For zero temperature the solution is drastically simplified. $P ( h )$ has the
form of a series of delta-function peaks at integer values of $h$
\begin{equation}
   P ( h ) = \sum_k p_k \delta ( h - k ) .
\end{equation}
These magnetic fields can be interpreted as follows. Non-zero values of $h$
($|h| \geqslant 1$) correspond to {\em frozen} variables. Indeed $h_i
\geqslant 1$ corresponds to $m_i = + 1$ and $h_i \leqslant - 1$ corresponds to
$m_i = - 1$. The appearance of finite fraction of such frozen spins, or
backbone, signals the beginning of unsatisfiable phase. The absolute value of
$h_i$ ($|h_i |$) then indicates the increase in the number of violated clauses
if the frozen spin is flipped.

Since the problem is symmetric with respect to sign flip $s_i \rightarrow -
s_i$, $P ( h )$ is necessarily symmetric. Introduce variable $q$
\begin{equation}
   q = \sum_{k = 1}^{\infty} p_k
\end{equation}
Then obviously $p_0 = 1 - 2 q$ since all $p_k$'s must add up to $1$. The
self-consistency equation can be written in the following form
\begin{eqnarray}
  1 - 2 q & = & \mathe^{- 3 \gamma q^2} I_0 ( 3 \gamma q^2 )\\
  p_k & = & \mathe^{- 3 \gamma q^2} I_k ( 3 \gamma q^2 )
\end{eqnarray}
Equivalently, these equations can be derived as follows \cite{duxbury}.
Let the fraction of frozen variables be $2 q$, with half of these being
polarized to $+ 1$ and
another half to $- 1$. Next, we add $N + 1$-st (cavity) spin to the system
together with extra clauses. The number of extra clauses is Poisson-distributed
with parameter $3 \gamma$. Each extra clause involves a cavity
spin and two randomly chosen variables of $N$-spin system. The clause forces a
certain value for the cavity spin only if both variables other than a cavity
spin are frozen to a value that equals corresponding $J_{m i}$. The
probability of this occurrence is $q^2$. Every such clause gives a contribution
$u = J_{m 1}$ to the effective field $h$ of cavity spin. This contribution is
$+ 1$ or $- 1$ with probability of 50 \%  each. If the probability that the
clause contributes $u = + 1$ (or $u = - 1$) is $q^2 / 2$, the number of
clauses attached to cavity spin that contribute $u = + 1$ (or $u = - 1$) is
Poisson with parameter $\frac{3}{2} \gamma q^2$. Therefore, we can write the
probability that the effective field of the cavity spin is $k$:
\begin{equation}
   p_k = \sum_{l, m} \frac{( 3 \gamma q^2 / 2 )^l}{l!}  \frac{( 3 \gamma q^2 /
   2 )^m}{m!} \mathe^{- 3 \gamma q^2} \delta_{l - m, k} \label{eq:pk}
\end{equation}
The right-hand side is evaluated to be $\mathe^{- 3 \gamma q^2} I_k ( 3 \gamma
q^2 )$. When the effective field equals zero, the cavity spin is not frozen.
Since the properties of $N$-spin system and $N + 1$-spin system should not
differ in the thermodynamic limit, the probability of this is $1 - 2 q$. Thus
we obtain the self-consistency equation on $q$.
\begin{equation}
   1 - 2 q = \mathe^{- 3 \gamma q^2} I_0 ( 3 \gamma q^2 ) .
\end{equation}
This equation does not determine $q$ uniquely; when both trivial solution ($q
= 0$) and a non-trivial ($q \neq 0$) are present, the correct value of $q$ is
chosen by examining the expression (\ref{eq:F})
for the free energy and choosing a
solution that maximizes the free energy. Although a non-trivial solution
appears at $\gamma \approx 4.67$, it does not become stable until
$\gamma_c \approx 5.18$

At finite but small temperature this picture is modified as follows. A series
of delta-function peaks in $P ( h )$ are broadened and acquire a finite $O ( T
)$ width (similar broadening occurs in quantum case for small $\Gamma$,
see figure \ref{fig:PhG}).
The values of $q$ and integrated probability weights around integer
values of $h$ remain the same. Effective fields for spins that are not frozen
$h \approx 0$ acquire $O ( T )$ corrections so that the magnetization $m =
\tanh \beta h = O ( 1 )$ and spins in the backbone ($|h| \gtrsim 1$) acquire
$O ( T )$ correction because of clauses connected to the backbone (see figure
\ref{fig:backbone}).

Examination of the free energy also shows that the value of $\gamma_c ( T )$
increases with increasing temperature. The following phase diagram is
obtained.

\begin{figure}[!ht]
\includegraphics{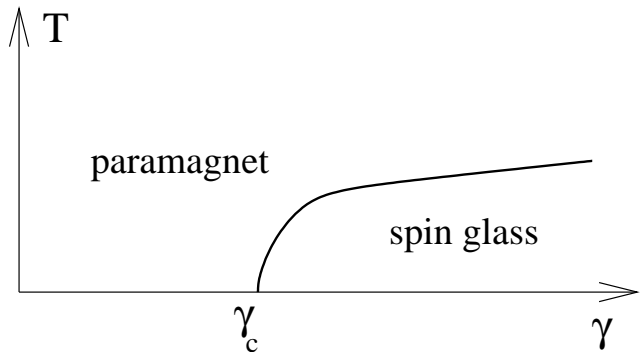}
\caption{Simple phase diagram obtained within RS ansatz}
\end{figure}

Unfortunately, the predicted value of $T = 0$ UNSAT transition of $\gamma_c
\approx \ldots .$ is far from the experimental value of $\gamma_c \approx
4.2$. The reason for the discrepancy is that the replica-symmetric ansatz
breaks down well before the satisfiability transition. Using an improved
1-step replica-symmetry breaking ansatz (1-step RSB) the break-down of replica
symmetry is put at $\gamma_d \approx 3.92$ and the satisfiability transition
is computed to occur at $\gamma_c \approx 4.27$ \cite{ksat2,ksat5}.

The central assumption of RSB ansatz is that for $\gamma > \gamma_d$ the
ground state (set of all satisfying assignments) is broken in an exponentially
large number of isolated islands separated by large barriers. The identity
$\langle s_i s_j \rangle = \langle s_i \rangle \langle s_j \rangle$ no longer
holds unless the thermal average is restricted to a particular local minimum
(or {\em pure state} in statistical physics lingo). Rather than being
described by individual magnetizations $m_i$, the system ought to be described
by a histogram of magnetizations taken over all possible pure states $\alpha$.
At the level if 1-step RSB, the following identity holds in the thermodynamic
limit. For sites $i$ and $j$ chosen at random (and as a result, typically
infinitely far away from each other) $P ( m_i m_j ) = P ( m_i ) P ( m_j )$,
where $P ( m_i )$ is a histogram of magnetizations at site $i$, and $P ( m_i,
m_j )$ is a histogram of pairs of magnetizations at sites $i$ and $j$.

Note that this ansatz is broken when once one goes beyond single step of
replica symmetry breaking. However, it is widely believed to be exact up to
$\gamma = \gamma_c$; thereafter, additional steps of RSB are required.

Due to the complex nature of the ground state in RSB phase, simulated
annealing algorithm should take an exponentially long time to converge to a
solution (although other specially designed algorithms can still work
efficiently for $\gamma > \gamma_d$). A correct phase diagram should take the
following form. The exponential slowing down that affects the performance of
simulated annealing algorithm occurs at RS/RSB boundary.

\begin{figure}[!ht]
\includegraphics{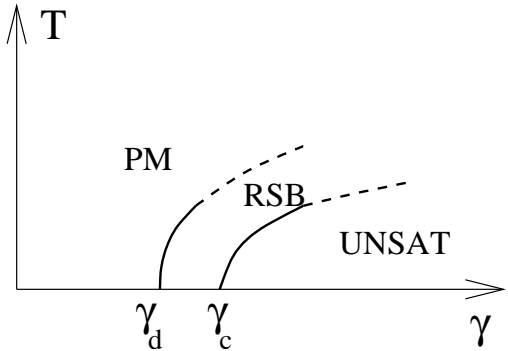}
\caption{Correct phase diagram obtained by considering effects of replica
symmetry breaking (RSB). RSB phase separate ``easy'' and unsatisfiable phases.}
\end{figure}

\section{\label{sec:q3sat}Quantum 3-SAT}

For quantum adiabatic evolution algorithm, the original Hamiltonian is
rewritten with operator $\hat{\sigma}_i^z$ replacing classical spins $s_i$ and
a driver term $- \Gamma \Sigma_i \hat{\sigma}_i^x$ is added
\begin{equation}
   \hat{H} = \hat{H}_{\textrm{cl}} [ \{ \hat{\sigma}_i^z \} ] - \Gamma \sum_i
   \hat{\sigma}_i^x .
\end{equation}
We expect level crossings and a critical slowing down to occur at phase
transition indicated by non-analyticities of quantum partition function
\begin{equation}
   Z = \textrm{Tr} \mathe^{- \beta \hat{H}} \label{eq:Z}
\end{equation}
Note that we will let $\beta \rightarrow \infty$ and examine the behavior of
$Z$ as a function of transverse field $\Gamma$.

To evaluate the partition function we use the Trotter lattice. Equation
(\ref{eq:Z}) can be rewritten as
\begin{equation}
   Z = \textrm{Tr} \left[ \mathe^{- ( \beta / M ) \hat{H}_{\textrm{cl}} [ \{
   \hat{\sigma}_i^z \} ] + ( \beta \Gamma / M ) \sum_i \hat{\sigma}_i^x}
   \right]^M  \label{eq:Z2}
\end{equation}
We can use Baker's identity
\begin{equation}
   \mathe^{\hat{A} + \hat{B}} = \mathe^{\hat{A}} \mathe^{\hat{B}} \mathe^{-
   \frac{1}{2} [ \hat{A}, \hat{B} ] + \cdots} \label{eq:baker}
\end{equation}
where additional terms involve higher-order commutators. In our case both
$\hat{A}$ and $\hat{B}$ are $O ( 1 / M )$ hence the last term in
(\ref{eq:baker}) is $\mathe^{O ( 1 / M^2 )}$ and does not contribute
to (\ref{eq:Z2}) in the limit $M
\rightarrow \infty$.
\begin{equation}
   Z = \textrm{Tr} \left[ \mathe^{- ( \beta / M ) \hat{H}_\textrm{cl}
   [ \{ \hat{\sigma}_i^z \} ]} \mathe^{( \beta \Gamma / M ) \sum_i
   \hat{\sigma}_i^x} \right]^M .
\end{equation}
To evaluate this expression for each of $M$ factors we insert basis states
corresponding to all $2^N$ projections of $\hat{\sigma}_i^z$. We denote basis
states by vector $s_{i, t}$ where $t = 0, \ldots, M$ labels a factor in which
the basis states appear. Since we are taking a trace we are imposing a
periodicity condition $s_{i, 0} = s_{i, M}$. Since $\hat{H}_{\textrm{cl}}$ is
obviously diagonal in this basis, first term is simply written as
\begin{equation}
   \mathe^{- ( \beta / M ) H_{\textrm{cl}} [ \{ s_{i, t} \} ]}
\end{equation}
and using $\mathe^{a \hat{\sigma}^x} = \textrm{ch} a + \hat{\sigma}^x \textrm{sh}
a$, second term can be written as
\begin{equation}
   \left( \tmop{ch} \frac{\beta \Gamma}{M} \right)^N \mathe^{\ln \tmop{th} (
   \beta \Gamma / M ) \sum_i ( 1 - s_{i, t} s_{i, t + 1} ) / 2} .
\end{equation}
The partition function of system with $N$ quantum spins is thus written
effectively as a partition function of classical system with $NM$ classical
spins.
\begin{equation}
   Z = ( \tmop{ch} \Gamma \Delta t )^{NM}  \sum_{\{ s_{i, t} \}} \mathe^{-
   \Delta t \sum_t H_{\tmop{cl}} [ \{ s_{i, t} \} ] + \ln \tmop{th} ( \Gamma
   \Delta t ) \sum_{i, t} ( 1 - s_{i, t} s_{i, t + 1} ) / 2},
\end{equation}
where we have denoted $\Delta t = \beta / M$ for readability. We are obviously
interested in the limit $M \rightarrow \infty$, $\beta \rightarrow \infty$,
$\Delta t \rightarrow 0$.

In what following we shall focus on the case with 3-bits in a
clause (3-SAT). As we have described earlier, the classical
Hamiltonian $H_{\tmop{cl}}$ for 3-SAT takes the following form:
\begin{equation}
   H_{\tmop{cl}} [ \{ s_{i, t} \} ] = 2 \sum_a \theta ( J_{a 1} s_{i_a} )
   \theta ( J_{a 2} s_{j_a} ) \theta ( J_{a 3} s_{k_a} )
\end{equation}
Variables $i_a, j_a$ and $k_a$ are drawn randomly out of $1, \ldots, N$;
variables $J_{a 1}, J_{a 2}, J_{a 3}$ are equal to $+ 1$ or $- 1$ with
probability of $1 / 2$. A particular realization of these variables is
referred to as disorder.

In replica method we prepare $n$ identical copies of the system so that the
partition function becomes $Z^n$. In practice, spin variables are augmented
with a replica index $\alpha = 1, \ldots, n$
\begin{equation}
   Z^n = ( \tmop{ch} \Gamma \Delta t )^{nNM}  \sum_{\{ s_{i, t}^{\alpha} \}}
   \mathe^{- \Delta t \sum_{\alpha, t} 2 \sum_a \theta ( J_{a 1} s_{i_a} )
   \theta ( J_{a 2} s_{j_a} ) \theta ( J_{a 3} s_{k_a} ) + \ln \tmop{th} (
   \Gamma \Delta t ) \sum_{\alpha, i, t} ( 1 - s_{i, t}^{\alpha} s_{i, t +
   1}^{\alpha} ) / 2} \label{eq:Zn}
\end{equation}
Note that after disorder average is taken, Hamiltonian becomes symmetric with
respect to permutations of sites labeled by index $i$ (but not to permutation
of replica indices $\alpha$ or Trotter indices $t$). Due to this permutation
symmetry, mean field theory should be exact. We introduce the following
variables
\begin{equation}
   c ( \{ \sigma_t^{\alpha} \} ) = \frac{1}{N}  \sum_i \prod_{\alpha, t}
   \delta [ s_{i, t}^{\alpha} ; \sigma_t^{\alpha} ] .
\end{equation}
This counts the fraction of sites for which spin assignment equals certain
vector $\sigma_t^{\alpha}$. By definition $\Sigma_{\{ \sigma_t^{\alpha} \}} c
( \{ \sigma_t^{\alpha} \} ) = 1$. Disorder-averaged summand of (\ref{eq:Zn})
can be expressed entirely in terms of $c ( \{ \sigma_t^{\alpha} \} )$. The
summand can be written as $\exp ( V_1 + V_2 )$, where the first term is due
to clauses
\begin{equation}
   V_1 [ c (\tmmathbf{\sigma}) ] = \gamma N \sum_{\{ \rho_t^{\alpha} \}, \{
   \sigma_t^{\alpha} \}, \{ \tau_t^{\alpha} \}} c (\tmmathbf{\rho}) c
   (\tmmathbf{\sigma}) c (\tmmathbf{\tau}) \left\langle \mathe^{- 2 \Delta t
   \sum_{\alpha, t} \theta ( J_1 \rho_t^{\alpha} ) \theta ( J_2
   \sigma_t^{\alpha} ) \theta ( J_3 \tau_t^{\alpha} )} \right\rangle
\end{equation}
the remaining average being only over the signs of $J_1, J_2, J_3$.

Second term is due to the transverse magnetic field
\begin{equation}
   V_2 [ c (\tmmathbf{\sigma}) ] = N \ln \tmop{th} ( \Gamma \Delta t )
   \sum_{\{ \sigma_t^{\alpha} \}} c (\tmmathbf{\sigma}) \sum_{\alpha, t}
   \frac{1 - \sigma_t^{\alpha} \sigma_{t + 1}^{\alpha}}{2}
\end{equation}
In the limit $N \rightarrow \infty$ sum over all possible realizations of $c
(\tmmathbf{\sigma})$ can be replaced by a continuous integral. The entropic
term (due to multiple realizations of $\{ s_{i, t}^{\alpha} \}$ with identical
$c (\tmmathbf{\sigma})$) takes a simple form
\begin{equation}
   S [ c (\tmmathbf{\sigma}) ] = - N \sum_{\tmmathbf{\sigma}} c
   (\tmmathbf{\sigma}) \ln c (\tmmathbf{\sigma}) .
\end{equation}
The partition function can be written as
\begin{equation}
   \left\langle Z^n \right\rangle = ( \tmop{ch} \Gamma \Delta t )^{nNM} \int [
   \mathd c (\tmmathbf{\sigma}) ] \mathe^{S [ c (\tmmathbf{\sigma}) ] + V_1 [
   c (\tmmathbf{\sigma}) ] + V_2 [ c (\tmmathbf{\sigma}) ]} .
\end{equation}
Since terms in the exponential are $O ( N )$, the value of the integral is
determined by its saddle-point.

\subsection{Replica-Symmetric Ansatz}

A major simplification is obtained if we assume that $c ( \{ \sigma_t^{\alpha}
\} )$ is symmetric under permutation of replica indices $\alpha$. For
simplicity, the following substitution is made
\begin{equation}
   c (\tmmathbf{\sigma}) = \int [ \mathd H (\tmmathbf{\tau}) ] P [ H
   (\tmmathbf{\tau}) ] \prod_{\alpha} \frac{\mathe^{H
   (\tmmathbf{\sigma}^{\alpha} )}}{\sum_{\{ \tau_t \}} \mathe^{H
   (\tmmathbf{\tau})}}
\end{equation}
where the integral is over all possible single-site Hamiltonians (i.e. all
possible vectors of size $2^M$ corresponding to all spin configurations on
Trotter lattice).

This has the desired symmetry property and is more amenable to taking $n
\rightarrow 0$ limit, since $c (\tmmathbf{\sigma})$ is now encoded by the
probability distribution $P [ H (\tmmathbf{\tau}) ]$. Maximization over all
possible $c (\tmmathbf{\sigma})$ is replaced by maximization over all possible
$P [ H (\tmmathbf{\tau}) ]$.

The entropic term is computed as follows
\begin{equation}
    - \sum_{\{ \sigma_t^{\alpha} \}} c (\tmmathbf{\sigma}) \ln c
   (\tmmathbf{\sigma}) = \left. - \frac{\mathd}{\mathd p} \sum_{\{ \sigma_t^{\alpha}
   \}} [ c (\tmmathbf{\sigma}) ]^p \right|_{p = 1}
\end{equation}
Substituting replica-symmetric ansatz for $c (\tmmathbf{\sigma})$ we obtain
\begin{equation}
   \sum_{\{ \sigma_t^{\alpha} \}} [ c (\tmmathbf{\sigma}) ]^p = \int [ \mathd
   H_1 (\tmmathbf{\sigma}) ] P [ H_1 (\tmmathbf{\sigma}) ] \ldots [ \mathd H_p
   (\tmmathbf{\sigma}) ] P [ H_p (\tmmathbf{\sigma}) ] \left( \frac{Z [ H_1
   (\tmmathbf{\sigma}) + \ldots + H_p (\tmmathbf{\sigma}) ]}{Z [ H_1
   (\tmmathbf{\sigma}) ] \ldots Z [ H_p (\tmmathbf{\sigma}) ]} \right)^n,
\end{equation}
where we have denoted $Z [ H (\tmmathbf{\sigma}) ] = \sum_{\{ \sigma_t \}}
\mathe^{H (\tmmathbf{\sigma})}$.

Taking the limit $n \rightarrow 0$ we use the fact $x^n \approx 1 + n \ln x$
and keep only contribution linear in $n$:
\begin{eqnarray}
  \sum_{\{ \sigma_t^{\alpha} \}} [ c (\tmmathbf{\sigma}) ]^p & = & n \int [
  \mathd H_1 (\tmmathbf{\sigma}) ] P [ H_1 (\tmmathbf{\sigma}) ] \ldots [
  \mathd H_p (\tmmathbf{\sigma}) ] P [ H_p (\tmmathbf{\sigma}) ] \ln Z [ H_1
  (\tmmathbf{\sigma}) + \ldots + H_p (\tmmathbf{\sigma}) ]\\
  &  & - np \int [ \mathd H (\tmmathbf{\sigma}) ] P [ H (\tmmathbf{\sigma}) ]
  \ln Z [ H (\tmmathbf{\sigma}) ]
\end{eqnarray}
Note that $H (\tmmathbf{\sigma})$ is defined only up to a constant. Without
losing generality we can require that $\sum_{\{ \sigma_t \}} H
(\tmmathbf{\sigma}) = 0$. With this constraint, we define the Fourier
transform of $P [ H (\tmmathbf{\sigma}) ]$:
\begin{equation}
   \tilde{P} [ \omega (\tmmathbf{\sigma}) ] = \int [ \mathd H
   (\tmmathbf{\sigma}) ] \mathe^{- i \sum_{\{ \sigma_t \}} \omega
   (\tmmathbf{\sigma}) H (\tmmathbf{\sigma})}
\end{equation}
Owing to the constraint $\sum_{\{ \sigma_t \}} H (\tmmathbf{\sigma}) = 0$,
Fourier transform $\tilde{P} [ \omega (\tmmathbf{\sigma}) ]$ is invariant
under shift $\omega (\tmmathbf{\sigma}) \rightarrow \omega (\tmmathbf{\sigma})
+ \omega_0$. With the aid of this Fourier transform the first term (which is
basically the integral of convolution of $P [ H (\tmmathbf{\sigma}) ]$) can be
rewritten as
\begin{equation}
   \int [ \mathd H (\tmmathbf{\sigma}) ] \ln Z [ H (\tmmathbf{\sigma}) ] \int
   [ \mathd \omega (\tmmathbf{\sigma}) ] \tilde{P} [ \omega
   (\tmmathbf{\sigma}) ]^p \mathe^{i \sum_{\{ \sigma_t \}} \omega
   (\tmmathbf{\sigma}) H (\tmmathbf{\sigma})}
\end{equation}
(Note: a factor of $( 2 \pi )^{- 2^M}$ is implicit in $[ \mathd \omega
(\tmmathbf{\sigma}) ]$)

This makes differentiation over $p$ trivial. The final result is
\begin{equation}
   S = - N \int [ \mathd H (\tmmathbf{\sigma}) ] \ln Z [ H (\tmmathbf{\sigma})
   ] \int [ \mathd \omega (\tmmathbf{\sigma}) ] \tilde{P} [ \omega
   (\tmmathbf{\sigma}) ] ( \ln \tilde{P} [ \omega (\tmmathbf{\sigma}) ] - 1 )
   \mathe^{i \sum_{\{ \sigma_t \}} \omega (\tmmathbf{\sigma}) H
   (\tmmathbf{\sigma})}
\end{equation}
The remaining contributions are obtained straightforwardly
\begin{equation}
   V_1 = \gamma N \int \left[ \left\langle \ln Z_3 [ H_1 (\tmmathbf{\sigma}),
   H_2 (\tmmathbf{\sigma}), H_3 (\tmmathbf{\sigma}) ] \right\rangle_{J_1, J_2,
   J_3} - \sum_{i = 1, 2, 3} \ln Z [ H_i (\tmmathbf{\sigma}) ] \right]
   \prod_{i = 1, 2, 3} P [ H_i (\tmmathbf{\sigma}) ] [ \mathd H_i
   (\tmmathbf{\sigma}) ],
\end{equation}
where the averaging is done over the signs of $J_1, J_2, J_3$ and $Z_3$ is
defined as follows
\begin{equation}
   Z_3 [ H_1 (\tmmathbf{\sigma}), H_2 (\tmmathbf{\sigma}), H_3
   (\tmmathbf{\sigma}) ] = \sum_{\{ \rho_t \}, \{ \sigma_t \}, \{ \tau_t \}}
   \mathe^{H_1 (\tmmathbf{\rho}) + H_2 (\tmmathbf{\sigma}) + H_3
   (\tmmathbf{\tau}) - 2 \Delta t \sum_t \theta ( J_1 \rho_t ) \theta ( J_2
   \sigma_t ) \theta ( J_3 \tau_t )}.
\end{equation}
And for $V_2$ we obtain the following expression
\begin{equation}
   V_2 = N \int [ \mathd H (\tmmathbf{\sigma}) ] P [ H (\tmmathbf{\sigma}) ]
   \left[ M \ln \tmop{ch} ( \Gamma \Delta t ) + \ln \sum_{\{ \sigma_t \}}
   \mathe^{H (\tmmathbf{\sigma}) + \ln \tmop{th} ( \Gamma \Delta t ) \sum_t (
   1 - \sigma_t \sigma_{t + 1} ) / 2} - \ln Z [ H (\tmmathbf{\sigma}) ]
   \right] .
\end{equation}

\subsection{Static Approximation}

Since working with the most general form of $H (\tmmathbf{\sigma})$ is
intractable, we make an ansatz
\begin{equation}
   H (\tmmathbf{\sigma}) = h \Delta t \sum_t \sigma_t + \ln \tmop{th} ( \Gamma
   \Delta t ) \sum_t ( 1 - \sigma_t \sigma_{t + 1} ) / 2
\end{equation}
with a single parameter $h$. This describes an isolated spin subject to
time-independent external magnetic field $h$. Alas, any dynamic effects are
neglected within this approximation. This is similar in spirit to the static
approximation made in solving infinitely-connected model in \cite{qsg1}.
Neglecting dynamic effects still permitted to obtain a qualitative picture of
the spin glass phase. It is widely believed that static approximation works
best in the limit of small $\Gamma$. We specifically consider that limit in
the next section.

The important simplification of present approach is that the order parameter
becomes a simple function $P ( h )$. Moreover, we can now take a limit $M
\rightarrow \infty$ further simplifying calculations. To the lowest order in
$\Delta t$
\begin{eqnarray}
  S & = & - N \int^{\infty}_{-\infty} \mathd h \ln [ 2 \tmop{ch} \beta \sqrt{\Gamma^2 + h^2} ]
  \int^{\infty}_{-\infty} \frac{\mathd \omega}{2 \pi} \mathe^{i \omega h}  \tilde{P} ( \omega ) [
  \ln \tilde{P} ( \omega ) - 1 ]\\
  &  & - N \ln ( \Gamma \Delta t ) \int^{\infty}_{-\infty} \mathd hP ( h ) \,\beta \Gamma
  \tmop{th}\left( \beta \sqrt{\Gamma^2 + h^2}\right),
\end{eqnarray}
where $\tilde{P} ( \omega )$ is the Fourier transform of $P ( h )$.

For $V_1$ we obtain the following expression
\begin{equation}
   V_1 = \gamma N \int \left[ \left\langle \ln Z_3 ( h_1, h_2, h_3 )
   \right\rangle_{J_1, J_2, J_3} - \sum_{i = 1, 2, 3} \ln Z_1 ( h_i ) \right]
   \prod_{i = 1, 2, 3} P ( h_i ) \mathd h_i,
\end{equation}
where $Z_1$ and $Z_3$ can, respectively, be expressed as follows
\begin{eqnarray}
  Z_1 ( h ) & = & \tmop{Tr} \mathe^{\beta ( h \hat{\sigma}_z + \Gamma
  \hat{\sigma}_x )},\\
  Z_3 ( h_1, h_2, h_3 ) & = & \tmop{Tr} \mathe^{\beta \sum_{i = 1, 2, 3} ( h_i
  \hat{\sigma}_{i z} + \Gamma \hat{\sigma}_{i x} ) - 2 \beta \theta ( J_1
  \hat{\sigma}_{1 z} ) \theta ( J_2  \hat{\sigma}_{2 z} ) \theta ( J_3
  \hat{\sigma}_{3 z} )} .
\end{eqnarray}
For $V_2$ the following expression is obtained to leading order in $\Delta t$.
\begin{equation}
   V_2 = N \ln ( \Gamma \Delta t ) \int^{\infty}_{-\infty} \mathd hP ( h ) \beta \Gamma \tmop{th}
   \beta \sqrt{\Gamma^2 + h^2}
\end{equation}
which exactly cancels $\Delta t$ dependence in $S$. Hence, as expected, for
sufficiently large $M$, solution is independent of $M$.

Observe that correct classical free energy is obtained if $\Gamma$ is set to
zero for finite $\beta$. For purely quantum result we set $\beta = + \infty$,
so that the free energy $F = - ( S + V_1 + V_2 ) / \beta$ becomes
\begin{eqnarray}
\frac{F}{N} & = &   \int^{\infty}_{-\infty} \mathd h
\sqrt{\Gamma^2 + h^2} \int^{\infty}_{-\infty} \frac{\mathd
  \omega}{2 \pi} \mathe^{i \omega h}  \tilde{P} ( \omega ) [ \ln \tilde{P} (
  \omega ) - 1 ]\\
  &  & - \gamma  \int \left[ \left\langle \Lambda ( J_! h_1, J_2 h_2, J_3
  h_3 ) \right\rangle_{J_1, J_2, J_3} - \sum_{i = 1, 2, 3} \sqrt{\Gamma^2 +
  h_i^2} \right] \prod_{i = 1, 2, 3} P ( h_i ) \mathd h_i \label{eq:F2}
\end{eqnarray}
where
\begin{equation}
   \Lambda ( J_1 h_1, J_2 h_2, J_3 h_3 ) = \lambda_{\max} \left[ \sum_{i = 1,
   2, 3} h_i  \hat{\sigma}_{i z} + \Gamma \sum_{i = 1, 2, 3} \hat{\sigma}_{i
   x} - 2 \prod_{i = 1, 2, 3} \theta ( J_i  \hat{\sigma}_{i z} ) \right],
\end{equation}
with $\lambda_{\max}$ denoting the largest eigenvalue of corresponding $8
\times 8$ matrix. The result depends on $J_1, J_2, J_3$ only through products
$J_! h_1, J_2 h_2, J_3 h_3$. It is subsequently averaged over signs of $J_1$,
$J_2$, $J_3$.

Differentiating the free energy with respect to $P ( h )$ allows us to write
self-consistency equation for $P ( h )$.
\begin{equation}
   \tilde{P} ( \omega ) = \exp 3 \gamma \frac{\int \left[ \left\langle \Lambda
   ( J_1 h_1, J_2 h_2, J_3 h_3 ) \right\rangle_{J_1, J_2, J_3} - \sum_{i = 1,
   2, 3} \sqrt{\Gamma^2 + h_i^2} \right] \mathe^{i \omega h_1} \mathd h_1 P (
   h_2 ) \mathd h_2 P ( h_3 ) \mathd h_3}{\int^{\infty}_{-\infty} \sqrt{\Gamma^2 + h^2} \mathe^{i
   \omega h} \mathd h}
\end{equation}
Replacing $J_i h_i$ by $h_i$ the averaging can be thrown out
\begin{equation}
   \tilde{P} ( \omega ) = \exp 3 \gamma \frac{\int \Lambda' ( h_1, h_2, h_3 )
   \cos \omega h_1 \mathd h_1  \frac{P ( h_2 ) + P ( - h_2 )}{2} \mathd h_2
   \frac{P ( h_3 ) + P ( - h_3 )}{2} \mathd h_3}{\int^{\infty}_{-\infty} \sqrt{\Gamma^2 + h^2}
   \cos \omega h \mathd h},
\end{equation}
where $\Lambda' ( h_1, h_2, h_3 ) = \Lambda ( h_1, h_2, h_3 ) - \sum_{i = 1,
2, 3} \sqrt{\Gamma^2 + h_i^2}$.

From this form it is also evident that $\tilde{P} ( \omega )$ is real and
hence $P ( h )$ is symmetric $P ( h ) = P ( - h )$, as it should be by the
symmetry of the model.

In high-$\Gamma$, low-connectivity phase we expect a solution to the
self-consistency equation to be unique, whereas in high-connectivity,
low-$\Gamma$ phase two solutions are present. To determine which solution to
take, one must examine the free energy $F$ and choose a solution that
maximizes its value. By rewriting (\ref{eq:F2}) with the aid of
self-consistency equation a somewhat simpler expression is obtained
\begin{eqnarray}
  \frac{F}{N} & = & 2 \gamma \int \Lambda' ( h_1, h_2, h_3 ) \prod_{i = 1, 2, 3} P (
  h_i ) \mathd h_i\\
  &  & - 3 \gamma \int \left[ \frac{\Lambda' ( + \infty, h_2, h_3 ) +
  \Lambda' ( - \infty, h_2, h_3 )}{2} \right] \prod_{i = 2, 3} P ( h_i )
  \mathd h_i - \int^{\infty}_{-\infty} \sqrt{\Gamma^2 + h^2} P ( h ) \mathd h \nonumber
  \label{eq:F3}
\end{eqnarray}

\section{\label{sec:small} Small $\Gamma$ limit}

\begin{figure}[!ht]
\includegraphics[scale=0.5]{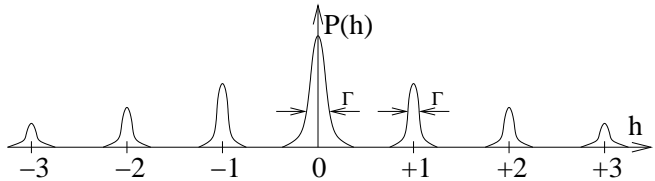}
\caption{Probability distribution of effective fields for small
transverse field $\Gamma$. \label{fig:PhG}}
\end{figure}

We examine the solution to the self-consistency equation in the
limit of small $\Gamma$. We seek a solution in the form of a
series of peaks of width $O ( \Gamma )$ centered at integer values
of $h$ (see Fig.~\ref{fig:PhG}). This behavior is similar similar
to what happens in classical case for small temperatures.

We can use the degenerate perturbation theory to evaluate $\Lambda
( h_1, h_2, h_3 )$. It will invariably involve finding the maximum
eigenvalue of some matrix corresponding to nearly degenerate
levels. To facilitate the discussion we denote by $L_2(a)$,
$L_3(a,b)$, $L_4(a,b,c)$, $L_7(a,b,c)$, the largest eigenvalues of
certain  matrices ${\cal M}_j$ ($j=2,3,4,7$) given in Appendix.

\begin{figure}[!ht]
\includegraphics[scale=0.75]{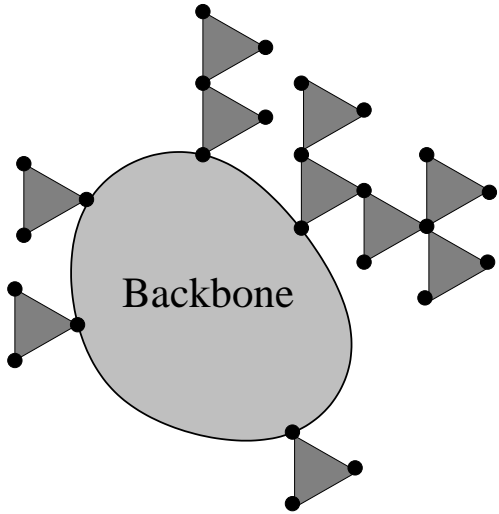}
\caption{Spins in the backbone (black area) together with some clauses
connected to them. The effect of the clauses is $O(\Gamma)$ corrections to
effective fields of spins in the backbone. \label{fig:backbone}}
\end{figure}

The reason for finite width of peaks is very similar to that for
classical case. Variables in isolated clusters and variables in
the backbone connected to free variables attain a small $O (
\Gamma )$ to their effective fields (see figure
\ref{fig:backbone}). Since trees of arbitrary depth are
technically possible, exact solution requires exact
diagonalization of arbitrary large matrices. The nature of the
static approximation that we made replaces effects of distant
clauses by an effective field; due to this the maximum size of the
matrix to be diagonalized is $2^3 \times 2^3$ -- the same as for
isolated clauses. Since we are doing this procedure in
self-consistent manner, it is better than simply truncating the
expansion in size of the cluster.

The matrices we consider above are submatrices of $8 \times 8$ matrix that
correspond only to those rows and columns that involve only combinations
of three spins that keep clause satisfied. For isolated clause this leaves
only 7 combinations. In the limit of $\Gamma=0$ these correspond to the 7
degenerate energy levels; finite $\Gamma$ lifts the the degeneracy and
diagonalization of $7 \times 7$ matrix is required. When one or more spins
in a clause are frozen, the degeneracy is smaller. The above-mentioned
matrices enumerate all possible degeneracies for $K=3$. All possible
combinations of effective fields that give rise to thses expressions are given
below.

With the aid of the notation that we just introduced, we will be able to write
the resulting expressions in compact form. We focus on the interaction term in
the expression for the free energy
\begin{equation}
   V = N \gamma \int \left[ \Lambda ( h_1, h_2, h_3 ) - \sum_{i = 1, 2, 3}
   \sqrt{\Gamma^2 + h_i^2} \right]  \prod_{i = 1, 2, 3} P ( h_i ) \mathd h_i
\end{equation}
For the remainder of this section when we say $h \approx k$ where $k$ is an
integer, we actually mean that $( h - k ) / \Gamma = O ( 1 )$.

Let us evaluate the expression in parentheses $\Lambda' = \Lambda
( h_1, h_2, h_3 ) - \sum_{i = 1, 2, 3} \sqrt{\Gamma^2 + h_i^2}$.
Since the expression is symmetric we can assume $h_1 \leqslant h_2
\leqslant h_3$ without losing generality. To first order in
$\Gamma$ the following expressions are obtained for all possible
cases:
\begin{itemize}
  \item[1.] $h_1 < 0, |h_1 | \gg \Gamma$. In this case $\Lambda' = 0$.

  \item[2.] Neither $h_1 \approx 0$ nor $h_1 \approx 1$. Classical
  expression holds in this case. $\Lambda' = - 2 \min ( 1, h_1 )$.

  \item[3.] $h_1 \approx 0 \tmop{and} h_2 \gg \Gamma$. No degeneracy. Simple
  answer $\Lambda' = - h_1 - \sqrt{\Gamma^2 + h_1^2}$.

  \item[4.] $h_1 \approx 0, h_2 \approx 0 \tmop{and} h_3 \gg \Gamma$. Triple
  degeneracy. For $\Lambda'$ we obtain the following expression:
  \begin{equation} \Lambda' = \Gamma L_3 ( - \tilde{h}_1, - \tilde{h}_2 ) - \Gamma \sum_{i =
     1, 2} \sqrt{1 + \tilde{h}_i^2}, \end{equation}
  where $\tilde{h}_i = h_i / \Gamma$.

  \item[5.] $h_1 \approx 0, h_2 \approx 0, h_3 \approx 0$. 7-fold degeneracy. For
  $\Lambda'$ we have
  \begin{equation} \Lambda' = \Gamma L_7 ( - \tilde{h}_1, - \tilde{h}_2, - \tilde{h}_3 ) -
     \Gamma \sum_{i = 1, 2, 3} \sqrt{1 + \tilde{h}_i^2}, \end{equation}
  where $\tilde{h}_i = h_i / \Gamma$.

  \item[6.] $h_1 \approx 1 \tmop{and} ( h_2 - 1 ) \gg \Gamma$. Double degeneracy.
  $\Lambda' = - 2 + \Gamma L_2 ( \tilde{h}_1 ) - \Gamma \tilde{h}_1$.

  \item[7.] $h_1 \approx 1, h_2 \approx 1 \tmop{and} ( h_3 - 1 ) \gg
  \Gamma$.
  Triple degeneracy.
  \begin{equation}
     \Lambda' = - 2 + \Gamma L_3 ( \tilde{h}_1, \tilde{h}_2 ) - \Gamma (
     \tilde{h}_1 + \tilde{h}_2 ),
  \end{equation}
  where $\tilde{h}_i = ( h_i - 1 ) / \Gamma$.

  \item[8.] $h_1 \approx 1, h_2 \approx 1, h_3 \approx 1$. 4-fold degeneracy.
  \begin{equation}
     \Lambda' = - 2 + \Gamma L_4 ( \tilde{h}_1, \tilde{h}_2, \tilde{h}_3 ) -
     \Gamma ( \tilde{h}_1 + \tilde{h}_2 + \tilde{h}_3 ),
  \end{equation}
  where $\tilde{h}_i = ( h_i - 1 ) / \Gamma$.
\end{itemize}
The self-consistency equation for $P ( h )$ can be written in the following
form
\begin{equation}
   \tilde{P} ( \omega ) = \exp \frac{\int_{-\infty}^{\infty} \left( \frac{\delta V / N}{\delta P
   ( h )} \right) \mathe^{i \omega h} \mathd h}{K ( \omega )},
\end{equation}
where $K ( \omega ) = \int \sqrt{\Gamma^2 + h^2} \mathe^{i \omega h} \mathd
h$. It is convenient to write
\begin{equation}
   K ( \omega ) = - \frac{2}{\omega^2} Q ( \Gamma \omega ),
\end{equation}
where $Q ( x )$ can be expressed using modified Bessel function of the second
kind
\begin{equation}
   Q ( x ) = \frac{|x|}{\sqrt{2 \pi}} K_1 ( |x| ) .
\end{equation}
Also observe that $Q ( 0 ) = 1$. With this replacement we can write
\begin{equation}
   \tilde{P} ( \omega ) = \exp \frac{1}{Q ( \omega )}  \int_{-\infty}^{\infty} \left[ \frac{1}{2}
   \frac{\mathd^2}{\mathd h^2} \left( \frac{\delta V / N}{\delta P ( h )}
   \right) \right] \mathe^{i \omega h} \mathd h.
\end{equation}
We seek a solution in the form of a sequence of peaks around integer values of
$h$ ($h \approx k$), each peak having a width $O ( \Gamma )$. Write
\begin{equation}
   P ( h ) = \frac{1}{\Gamma} \sum_{k = - \infty}^{+ \infty} P_k \left(
   \frac{h - k}{\Gamma} \right) .
\end{equation}
Total probability weight around $h \approx k$ can be expressed as
$p_k = \int_{-\infty}^{\infty} P_k ( \tilde{h} ) \mathd
\tilde{h}$.

The interaction term $V$ has the form proportional to
\begin{equation}
   \int \Lambda' ( h_1, h_2, h_3 ) P ( h_1 ) P ( h_2 ) P ( h_3 ) \mathd h_1
   \mathd h_2 \mathd h_3 .
\end{equation}
Computing a variation $\delta V / \delta P ( h )$ leaves only two
probability distributions in the integral; say $P ( h_2 )$ and $P
( h_3 )$.

First, we assume that neither $h \approx 0$ nor $h \approx 1$.
Note that the two remaining fields in the integrand: $h_2$ and $h_3$ are
necessarily integers.
If $h < h_2, h_3$ we know that $\Lambda' = \theta ( h ) \min ( 1, h )$, which,
differentiated twice over $h$ gives just zero.

If $h > h_2, h_3$ and $h - h_2 \gg \Gamma$, $h - h_3 \gg \Gamma$, then $\delta
V / \delta P ( h )$ is actually independent of $h$ and does not survive double
differentiation over $h$ either. Note that this holds even if $h \approx 1$
(but either $h_2 \approx 0$ or $h_3 \approx 0$).

Therefore, the only non-zero contributions are from $h \approx 0$ and $h
\approx 1$. Moreover expressions of the form $\int_{0 + \Gamma \infty}^{+
\infty} P ( h ) \mathd h$ and $\int_{1 + \Gamma \infty}^{+ \infty} P ( h )
\mathd h$ can be replaced with constants $q$ and $q - p_1$ ($q = \sum_{k =
1}^{+ \infty} p_k$) since we have seen already that varying with respect to $P
( h )$ in these expressions leads to $h$-independent terms.

It is convenient to rewrite $V$ in terms of $P_k ( \tilde{h} )$. We write
\begin{equation}
   V / N = 2 \gamma q^3 + \gamma \Gamma V',
\end{equation}
where $V'$ takes more complex form
\begin{eqnarray}
  V' & = & \int L_7 ( - \tilde{h}_1, - \tilde{h}_2, - \tilde{h}_3 ) \prod_{i =
  1, 2, 3} P_0 ( \tilde{h}_i ) \mathd \tilde{h}_i + 3 q \int L_3 ( -
  \tilde{h}_1, - \tilde{h}_2 ) \prod_{i = 1, 2} P_0 ( \tilde{h}_i ) \mathd
  h_i \nonumber\\
  &  & + 3 q^2 \int \tilde{h}_1 P_0 ( \tilde{h}_1 ) \mathd \tilde{h}_1 - 3 (
  p_0 + q )^2  \int \sqrt{1 + \tilde{h}^2} P_0 ( \tilde{h} ) \mathd
  \tilde{h} \nonumber\\
  &  & + \int L_4 ( \tilde{h}_1, \tilde{h}_2, \tilde{h}_3 ) \prod_{i = 1, 2,
  3} P_1 ( \tilde{h}_i ) \mathd \tilde{h}_i + 3 ( q - p_1 ) \int L_3 (
  \tilde{h}_1, \tilde{h}_2 ) \prod_{i = 1, 2} P_1 ( \tilde{h}_i ) \mathd
  \tilde{h}_i \nonumber\\
  &  & + 3 ( q - p_1 )^2 \int L_2 ( \tilde{h}_1 ) P_1 ( \tilde{h}_1 ) \mathd
  \tilde{h}_1 - 3 q^2 \int \tilde{h} P_1 ( \tilde{h} ) \mathd \tilde{h} .
  \label{eq:V1}
\end{eqnarray}
We can use the identity $\delta V / \delta P ( h ) = \delta V / \delta P_0 (
\tilde{h} )$ for $h \approx 0$ and $\delta V / \delta P ( h ) = \delta V /
\delta P_1 ( \tilde{h} )$ for $h \approx 1$. We have shown that all other
terms do not contribute.

For the partial derivatives we obtain the following expression (also taking
into account the fact that $P_0 ( \tilde{h} ) = P_0 ( - \tilde{h} )$.
\begin{eqnarray}
  \frac{1}{3 \gamma \Gamma}  \frac{\delta V / N}{\delta P_0 ( \tilde{h} )} & =
  & \int L_7 ( \tilde{h}, \tilde{h}_2, \tilde{h}_3 ) P_0 ( \tilde{h}_2 ) P_0 (
  \tilde{h}_3 ) \mathd \tilde{h}_2 \mathd \tilde{h}_3 + 2 q \int L_3 (
  \tilde{h}, \tilde{h}_2 ) P_0 ( \tilde{h}_2 ) \mathd \tilde{h}_2 \nonumber\\
  &  & + q^2  \tilde{h} - ( p_0 + q )^2 L_2 ( \tilde{h} )\\
  \frac{1}{3 \gamma \Gamma}  \frac{\delta V / N}{\delta P_1 ( \tilde{h} )} & =
  & \int L_4 ( \tilde{h}, \tilde{h}_2, \tilde{h}_3 ) P_1 ( \tilde{h}_2 ) P_1 (
  \tilde{h}_3 ) \mathd \tilde{h}_2 \mathd \tilde{h}_3 + 2 ( q - p_1 ) \int L_3
  ( \tilde{h}, \tilde{h}_2 ) P_1 ( \tilde{h}_2 ) \mathd \tilde{h}_2 \nonumber\\
  &  & + ( q - p_1 )^2 L_2 ( \tilde{h} ) - q^2  \tilde{h}
\end{eqnarray}
We write the self-consistency equation in the following form
\begin{equation}
   \tilde{P} ( \omega ) = \exp \frac{3 \gamma}{Q ( \omega )} \left[ F_0 (
   \Gamma \omega ) + \frac{1}{2} \left( \mathe^{i \omega} F_1 ( \Gamma \omega
   ) + \mathe^{- i \omega} F_1^{\ast} ( \Gamma \omega ) \right) \right]
\end{equation}
where we have denoted
\begin{eqnarray}
  F_0 ( \tilde{\omega} ) & = & \int^{\infty}_{-\infty} \left[ \frac{1}{2}
  \frac{\partial^2}{\partial \tilde{h}^2} \left( \frac{1}{3 \gamma \Gamma}
  \frac{\delta V / N}{\delta P_0 ( \tilde{h} )} \right) \right] \mathe^{- i
  \tilde{\omega}  \tilde{h}} \mathd \tilde{h},\\
  F_1 ( \tilde{\omega} ) & = & \int^{\infty}_{-\infty} \left[ \frac{1}{2}
  \frac{\partial^2}{\partial \tilde{h}^2} \left( \frac{1}{3 \gamma \Gamma}
  \frac{\delta V / N}{\delta P_1 ( \tilde{h} )} \right) \right] \mathe^{- i
  \tilde{\omega}  \tilde{h}} \mathd \tilde{h} .
\end{eqnarray}
Since for all functions $L_n$ we have the identity $\frac{1}{2}
\frac{\partial}{\partial \tilde{h}} L ( \tilde{h}, \ldots ) |^{\tilde{h} = +
\infty}_{\tilde{h} = - \infty} = 1$, we obtain $F_0 ( 0 ) = - q^2$ and $F_1 (
0 ) = q^2$.

We now perform the Fourier transform of $\tilde{P} ( \omega )$. Since
$\tilde{P} ( \omega )$ is modulated periodic function with $\omega_{\max} \sim
1 / \Gamma$, its Fourier transform is necessarily a series of spikes of width
$\Gamma$. This justifies our previous ansatz. Let us compute the probability
density for $h \approx k$.
\begin{equation}
   P_k ( \tilde{h} ) = \int_{-\infty}^{\infty} \mathe^{ik \omega + i \Gamma \omega \tilde{h}}
   \exp \frac{3 \gamma}{Q ( \omega )} \left[ F_0 ( \Gamma \omega ) + \tmop{Re}
   \left\{ \mathe^{i \omega} F_1 ( \Gamma \omega ) \right\} \right]
   \frac{\mathd \omega}{2 \pi} .
\end{equation}
Using the smallness of $\Gamma$ and the identity $\int_{- \pi}^{\pi} \mathe^{a
\cos \omega + ik \omega}  \frac{\mathd \omega}{2 \pi} = I_k ( a )$, we can
rewrite to leading order in $\Gamma$
\begin{equation}
   P_k ( \tilde{h} ) = \int^{\infty}_{-\infty} \mathe^{3 \gamma F_0 ( \tilde{\omega} ) / Q (
   \tilde{\omega} )} I_k \left( \frac{3 \gamma |F_1 ( \tilde{\omega} ) |}{Q (
   \tilde{\omega} )} \right) \mathe^{i \tilde{\omega}  \tilde{h}}
   \frac{\mathd \tilde{\omega}}{2 \pi} .
\end{equation}
Note that the integrated probability weights $p_k = \int P_k ( \tilde{h} )
\mathd \tilde{h}$ are obtained by substituting $\tilde{\omega} = 0$ in the
integrand
\begin{equation}
   p_k = \mathe^{- 3 \gamma q^2} I_k ( 3 \gamma q^2 ) .
\end{equation}
This is precisely Eq. \ref{eq:pk}.
For $k = 0$ using $p_0 = 1 - 2 q$ a self-consistency
equation identical to that of $\Gamma = 0$, $T = 0$ is obtained. Therefore,
the value of $q$ is unchanged to leading order in $\Gamma$.

Since $F_0 ( \tilde{\omega} )$ and $F_1 ( \tilde{\omega} )$ are given solely
in terms of $P_0 ( \tilde{h} )$ and $P_1 ( \tilde{h} )$, $k = 0$ and $k = 1$
are sufficient to provide a closed system of equations
\begin{eqnarray}
  P_0 ( \tilde{h} ) & = & \int^{\infty}_{-\infty} \mathe^{3 \gamma F_0 ( \tilde{\omega} ) / Q (
  \tilde{\omega} )} I_0 \left( \frac{3 \gamma |F_1 ( \tilde{\omega} ) |}{Q (
  \tilde{\omega} )} \right) \mathe^{i \tilde{\omega}  \tilde{h}}  \frac{\mathd
  \tilde{\omega}}{2 \pi},\\
  P_1 ( \tilde{h} ) & = & \int^{\infty}_{-\infty} \mathe^{3 \gamma F_0 ( \tilde{\omega} ) / Q (
  \tilde{\omega} )} I_1 \left( \frac{3 \gamma |F_1 ( \tilde{\omega} ) |}{Q (
  \tilde{\omega} )} \right) \mathe^{i \tilde{\omega}  \tilde{h}}  \frac{\mathd
  \tilde{\omega}}{2 \pi} .
\end{eqnarray}
For some critical value of $\gamma$ both trivial ($q = 0$) and a non-trivial
($q \neq 0$) solutions coexist. While the appearance of non-trivial $q$ and
its value are not sensitive to $\Gamma$ for small values of $\Gamma$, the
point where the non-trivial solution becomes stable is.

Also observe that since the integrand is real, all $P_k ( \tilde{h} )$ are
symmetric $P_k ( \tilde{h} ) = P_k ( - \tilde{h} )$. While this is expected
for $P_0 ( \tilde{h} )$, such symmetry for $P_1 ( \tilde{h} )$ and others is
likely only approximate, higher order contributions in $\Gamma$ should make
$P_k ( \tilde{h} ) \neq P_k ( - \tilde{h} )$ for $k \neq 0$.

Determining the stability of non-trivial solution is accomplished
with the aid of Eq.~(\ref{eq:F3}). Substituting our ansatz for $P
( h )$ we obtain
\begin{equation}
   \frac{F}{N}\equiv \frac{F(\Gamma)}{N} = 3 \gamma q^2 - 4 \gamma q^3 + \sum_k |k|p_k - 2 \gamma \Gamma V' -
   \frac{3}{2} \gamma \Gamma V'' + \Gamma \int \sqrt{1 + \tilde{h}^2} P_0 (
   \tilde{h} ) \mathd \tilde{h},
\end{equation}
where $V'$ has been defined in (\ref{eq:V1}) and $V''$
(coming from $\Lambda' ( +\infty, \ldots )$ term) can be written as
\begin{eqnarray}
  V'' & = & \int L_3 ( \tilde{h}_1, \tilde{h}_2 ) P_1 ( \tilde{h}_1 ) P_1 (
  \tilde{h}_2 ) \mathd \tilde{h}_1 \mathd \tilde{h}_2 \nonumber\\&+& 2 ( q - p_1 ) \int L_2
  ( \tilde{h}_1 ) P_1 ( \tilde{h}_1 ) \mathd h_1 - 2 q \int \tilde{h} P_1 (
  \tilde{h} ) \mathd \tilde{h} .
\end{eqnarray}
For $\Gamma = 0$ the last three terms disappear and substituting $p_k =
\mathe^{- 3 \gamma q^2} I_k ( 3 \gamma q^2 )$ we obtain
\begin{equation}
   \frac{F(0)}{N} = \gamma q^2 \left( 1 - \mathe^{- 3 \gamma q^2} \left( I_0 ( 3 \gamma
   q^2 ) + 3 I_1 ( 3 \gamma q^2 ) \right) \right)
\end{equation}\noindent
The value of $\gamma=\gamma_{c0} \approx 5.18$ where this
expression (with $q \neq 0$ determined from self-consistency
equation) becomes positive is the point where non-trivial solution
becomes stable. We now calculate the phase transition line
$\gamma=\gamma_c(\Gamma)$ to the leading order in $\Gamma$ (here
$\gamma(0)\equiv\gamma_{c0}$).
  For $\gamma$ slightly larger than $\gamma_{c0}$,
classical expression can be written as \begin{equation}
\frac{F(0)}{N} = A ( \gamma - \gamma_c ( 0 ) ),\qquad  A = \left .
\frac{\partial F(0,\gamma)}{\partial \gamma}\right|_{\gamma_c(0)}
\approx 0.0706.
\end{equation}
The remaining terms in $F / N$ are linear in $\Gamma$. Write them in the form
$- \Gamma \tilde{V} ( q )$:
\begin{eqnarray}
  \tilde{V} ( q ) & = & 2 \gamma \int L_7 ( \tilde{h}_1, \tilde{h}_2,
  \tilde{h}_3 ) \prod_{i = 1, 2, 3} P_0 ( \tilde{h}_i ) \mathd h_i + 6 \gamma
  q \int L_3 ( \tilde{h}_1, \tilde{h}_2 ) \prod_{i = 1, 2} P_0 ( \tilde{h}_i )
  \mathd \tilde{h}_i\\
  &  & + 2 \gamma \int L_4 ( \tilde{h}_1, \tilde{h}_2, \tilde{h}_3 ) \prod_{i
  = 1, 2, 3} P_1 ( \tilde{h}_i ) \mathd \tilde{h}_i + 6 \gamma \left(
  \frac{1}{4} + q - p_1 \right) \int L_3 ( \tilde{h}_1, \tilde{h}_2 ) \prod_{i
  = 1, 2} P_1 ( \tilde{h}_i ) \mathd \tilde{h}_i \nonumber\\
  &  & + 6 \gamma ( q - p_1 ) \left( \frac{1}{2} + q - p_1 \right) \int
  \sqrt{1 + \tilde{h}^2} P_1 ( \tilde{h} ) \mathd \tilde{h} - \left( 1 + 6
  \gamma ( 1 - q )^2 \right) \int \sqrt{1 + \tilde{h}^2} P_0 ( \tilde{h} )
  \mathd \tilde{h} \nonumber
\end{eqnarray}
Let $\tilde{V}_0 ( q_0 )$ denote $\tilde{V} ( q )$ computed at $\gamma =
\gamma_c ( 0 )$ with $P_0 ( \tilde{h} )$, $P_1 ( \tilde{h} )$ solving
self-consistency equation with $q \neq 0$; denote $\tilde{V}_0 ( 0 )$
similarly computed $\tilde{V} ( q )$ at $\gamma = \gamma_c$ for trivial
solution $q = 0$.

To leading order in $\Gamma$, the point $\gamma_c$ where nontrivial $q \neq 0$
solution becomes stable is
\begin{equation}
   \gamma_c ( \Gamma ) = \gamma_c ( 0 ) + \Gamma \frac{\tilde{V}_0 ( 0 ) -
   \widetilde{V_0} ( q_0 )}{A} .
\end{equation}

\section{\label{sec:conclusion} Conclusion}

In this paper we have extended the classical treatment of phase
transitions in K-SAT \cite{ksat1,ksat3,ksat4} to the quantum
domain for the case of $K=3$. Although infinitely-connected
quantum spin glass models have been studied, no studies of dilute
spin glasses have been performed to the best of our knowledge.

While infinitely-connected models have small ($O(1/\sqrt{N})$ or smaller)
couplings, dilute glasses are characterized by strong ($O(1)$) couplings
which means that perturbation expansion cannot be truncated. Due to the
limitations imposed by the structure of disorder, we have only been able to
solve the problem within static approximation using replica symmetric ansatz.

What we have observed is that the quantum limit $\Gamma > 0$,
$T=0$ is qualitatively similar to the classical limit $\Gamma=0$,
$T>0$ (although quantitative results and analytical expressions
are quite different). The order parameter in the limit of small
$\Gamma$ takes the form of series of peaks of width $O(\Gamma)$,
just as in the classical case it takes the form of series of peaks
of width $O(T)$. At $\Gamma=0$, $T=0$ we recovet the phase
transition at the classical value $\gamma_c(0)$. For small but
finite $\Gamma$ the value of $\gamma_c$ increases linearly,
$\gamma_c = \gamma_c(0)+C \Gamma$, and the expression for the
constant $C$ is given in the closed form in terms of the
quantities computed at $\Gamma=0$.

Much of the similarities can be explained away by the fact that we
ignored dynamic effects. Incorporating the dynamic effects changes
the phase diagram of Sherrington-Kirkpatrick model
\cite{dynamic,dynamic2,dynamic3}. However, static approximation is
assumed to work very well in the limit of small $\Gamma$. For the
problem at hand this seems to be the only feasible limit. A bigger
concern is that we have completely ignored the effects of replica
symmetry breaking (RSB). We expect that the location of the
dynamic transition for the QAA algorithm should be the same as for
simulated annealing, since it is given by $\Gamma=0$, $T=0$.
Working in the regime of small $T$ or small $\Gamma$ within RSB
will enable us to compare performance of these algorithms for
$\gamma \approx \gamma_d$.

Another suggestion for future work is K-XOR-SAT problem. It can be
solved on a classical computer in polynomial time, but becomes
exponentially hard for the simulated annealing algorithm. Owing to
simple structure of its energy landscape, many exact results have
been obtained for this problem \cite{kxorsat}. Quantum version of
$K$-XOR-SAT in the limit of small $\Gamma$ might be much easier to
solve. Note that many of properties of K-SAT are found in
K-XOR-SAT, example being a single level of replica symmetry
breaking in a certain range of $\gamma$.

We have worked within replica symmetric formalism developed by Monasson
\cite{monasson} that uses a functional order parameter. Alternative method
of working with dilute glasses that is closer in spirit to the treatment
of SK model has been developed by Viana and Bray \cite{vianabray} and it
incorporates a sequence of various order parameters. The bridge between
this approaches have been developed by Kanter and Sompolinsky
\cite{sompolinsky}. A more consistent approach to making static approximation
would be to ignore time dependence in a sequence of VB-like order parameters
along the lines of \cite{qsg1} and convert the result to the form
that uses functional order parameter. We have not tried to reconcile that
approach with our treatment. It is interesting to see if these approaches
are equivalent, and if not, how the answer changes.

\appendix
\section*{\label{appendix} Appendix} Matrices ${\cal M}_j$
introduced in the Sec.~\ref{sec:small} have the following form:
\begin{equation}
 {\cal M}_2=\left(
  \begin{array}{cc}
    a & 1\\
    1 & -a
  \end{array}
  \right),\end{equation}
\begin{equation}
{\cal  M}_3=\left(
  \begin{array}{ccc}
    a + b & 1 & 1\\
    1 & a - b & 0\\
    1 & 0 & - a + b
  \end{array}
  \right),\end{equation}
  \begin{equation}
{\cal  M}_4= \left(
  \begin{array}{cccc}
    a + b + c & 1 & 1 & 1\\
    1 & a + b - c & 0 & 0\\
    1 & 0 & a - b + c & 0\\
    1 & 0 & 0 & - a + b + c
  \end{array}
  \right),\end{equation}
  \begin{equation}
{\cal M}_7=\left(
  \begin{array}{ccccccc}
    a + b + c & 1 & 1 & 0 & 1 & 0 & 0\\
    1 & a + b - c & 0 & 1 & 0 & 1 & 0\\
    1 & 0 & a - b + c & 1 & 0 & 0 & 1\\
    0 & 1 & 1 & a - b - c & 0 & 0 & 0\\
    1 & 0 & 0 & 0 & - a + b + c & 1 & 1\\
    0 & 1 & 0 & 0 & 1 & - a + b - c & 0\\
    0 & 0 & 1 & 0 & 1 & 0 & - a - b + c
  \end{array}
  \right).
\end{equation}

\end{document}